\documentclass[epj,nopacs]{svjour}

\usepackage{graphics}
\usepackage[tbtags]{amsmath}

\begin{document}

\title{Birkhoff's Theorem in the $f(T)$ Gravity}
\author{Xin-He Meng\inst{1,2} \thanks{\email{xhm@nankai.edu.cn}}
\and Ying-Bin Wang \inst{1} \thanks{\email{wybn@mail.nankai.edu.cn}}
}

\institute{Department of physics, Nankai University, Tianjin 300071, China
\and Kavli Institute of Theoretical Physics China, CAS, Beijing 100190, China}

\date{Received: date / Revised version: date}

\abstract{ Generalized from the so-called teleparallel gravity which
is exactly equivalent to general relativity, the $f(T)$ gravity has
been proposed as an alternative gravity model to account for the
dark energy phenomena. In this letter we prove that the external
vacuum gravitational field for a spherically symmetric distribution
of source matter in the $f(T)$ gravity framework must be static and
the conclusion is independent of the radial distribution and
spherically symmetric motion of the source matter that is, whether
it is in motion or static. As a consequence, the Birkhoff's theorem
is valid in the general $f(T)$ theory. We also discuss its
application in the de Sitter space-time evolution phase as preferred
to by the nowadays dark energy observations.
\\\\
PACS.98.80.Cq Modified theories of gravity } \maketitle
\section{Introduction}

The discovery of the accelerating expansion of the Universe has
stimulated great efforts to investigate the fundamental theories of
gravity. As a modified gravitational theory, the $f(T)$ gravity has
been proposed \cite{fT} to explain the acceleration of the cosmic
expansion and attracts much attention recently. The framework is a
generalization of the so-called \textit{Teleparallel Equivalent of
General Relativity} (TEGR) which was first propounded by Einstein in
1928 \cite{einstein} and maturated in the 1960s (For some reviews,
see \cite{TEGR1,TEGR2}). We know that the theory of general
relativity is based on Riemann geometry which involves only
curvature composed of the metric and its derivatives. On the
contrary, the TEGR is based on the so named Weitzenb\"ock geometry
with the non-vanishing torsion. Owing to the definition of
Weitzenb\"ock connection rather than the Levi-Civita connection, the
Riemann curvature is automatically vanishing in the TEGR framework.
Therefore the parallelism of distant vectors or tensors would be
independent of curves along which they are transported. That is why
the theory is also called \textit{Teleparallel Gravity}. It has been
well studied that the TEGR, for a specific choice of parameters,
behaves completely equivalent to Einstein's theory of general
relativity. Furthermore, by using the torsion scalar $T$ as the
Lagrangian density, the TEGR can give a field equation of the second
order only, which is simpler than Einstein's field equation and
avoids the instability problems caused from higher order derivatives
as from the metric framework f(R) gravity demonstrated.

The modified version of teleparallel gravity uses a general function
$f(T)$ as the model Lagrangian density. Similar to the
generalization of Einstein's theory of general relativity to the
$f(R)$ theory (For some references, see \cite{fR1,fR2,fR3,fR4}), the
$f(T)$ theory can be directly reduced to the TEGR if we choose a
simplest case $f(T)=T$. A variety of $f(T)$ models have been
proposed in succession to explain the late-time acceleration of the
cosmic expansion without the mysteriously so-called dark energy, and
have been fitted the cosmological data-sets very well (e.g.
\cite{fT,fT1,fT2,fT3,fT4,fT5,fT6}). In the theoretical aspect, the
Lorentz invariance and conformal invariance of the $f(T)$ theory are
also investigated interestingly\cite{fT_Lorentz,fT_conformal}, and
present many interesting results. In this paper, we focus on the
validity of Birkhoff's theorem in the $f(T)$ gravity.

The Birkhoff's theorem, presented with an explicit proof by George
D. Birkhoff in 1923 \cite{birkhoff}, states that the spherically
symmetric gravitational field in vacuum must be static, with a
metric uniquely given by the Schwarzschild solution form of Einstein
equations \cite{weinberg}. It is well known that the Schwarzschild
metric is found in 1918 as the external (vacuum) solution of a
static and spherical star. The Birkhoff's theorem claims that any
spherically symmetric object possesses the same static gravitational
field, as if the mass of the object were concentrated at the center.
It is the same feature as holding in the classical Newtonian
gravity. It means that the external gravitational field is static
even if the central spherical object is moving radically, like the
collapsing processes (such as a collapsing star or a violently
exploding supernova), as long as the motion is spherically
symmetric. As a result, there is no monopole gravitational radiation
anyway, just as the case of electromagnetic radiation physics.

As a relatively newly proposed modified gravitational theory, we do
not know clearly for the $f(T)$ gravity whether the vacuum solution
of spherically symmetric field is still static before. In this
present paper, first we briefly review the $f(T)$ theories in the
following section, and in section three we prove the validity of
Birkhoff's theorem in the context of the $f(T)$ gravity. For the
sake of clarity, we firstly demonstrate the proof for a concrete
form of a $f(T)$ model, and then complete it for the general case.
The conclusions and discussions are devoted in the last section.

\section{Elements of $f(T)$ Gravity}

Instead of the metric tensor, the vierbein field
$\mathbf{e}_{i}(x^{\mu})$ is the dynamical variable in the
teleparallel gravity. It is defined as the orthonormal basis of the
tangent space at each point $x^{\mu}$ in the manifold, namely,
$\mathbf{e}_{i}\cdot \mathbf{e}_{j}=\eta_{ij}$, where
$\eta_{ij}=diag(1,-1,-1,-1)$ is the Minkowski metric. The vierbein
vector can be expanded in spacetime coordinate basis:
$\mathbf{e}_{i}=e^{\mu}_{i} \partial_{\mu}$,
$\mathbf{e}^i=e^i_\mu{\rm d}x^\mu$. According to the convention,
Latin indices and Greek indices, both running from 0 to 3, label the
tangent space coordinates and the spacetime coordinates
respectively. The components of vierbein are related by $e_{\mu}^i
e^{\mu}_j=\delta^{~i}_{j}$, ~~$e_{\mu}^i
e^{\nu}_i=\delta_{\mu}^{~\nu}$.

The metric tensor is determined uniquely by the vierbein as
\begin{equation}
g_{\mu\nu}=\eta_{ij} e_{\mu}^i e_{\nu}^i,
\end{equation}
which can be equivalently expressed as: $\eta_{ij}=g_{\mu\nu}
e^i_{\mu} e^j_{\nu}$. The definition of torsion tensor is given by
then
\begin{equation}
T^{\rho}_{~\mu\nu}=\Gamma^{\rho}_{~\nu\mu}-\Gamma^{\rho}_{~\mu\nu},
\end{equation}
where $\Gamma^{\rho}_{~\mu\nu}$ is the connection. Evidently,
$T^{\rho}_{~\mu\nu}$ vanishes in the Riemann geometry since the
Levi-Civita connection is symmetric with respect to the two
covariant indices. Differing from that in Einstein's theory of
general relativity, the teleparallel gravity uses Weitzenb\"ock
connection defined directly from the vierbein:
\begin{equation}
\Gamma^{\rho}_{~\mu\nu}=e_i^{\rho} \partial_{\nu} e^i_{\mu}.
\end{equation}
Accordingly the antisymmetric non-vanishing torsion is
\begin{equation}\label{torsion}
T^{\rho}_{~\mu\nu}=e_i^{\rho}(\partial_{\mu}e^i_{\nu} - \partial_{\nu}e^i_{\mu}).
\end{equation}
It can be confirmed that the Riemann curvature in this framework is precisely vanishing:
\begin{equation}
R^\rho_{~\theta\mu\nu}=\partial_\mu \Gamma^\rho_{~\theta\nu}-\partial_\nu \Gamma^\rho_{~\theta\mu}+\Gamma^\rho_
{~\sigma\mu}\Gamma^\sigma_{~\theta\nu}-\Gamma^\rho_{~\sigma\nu} \Gamma^\sigma_{\theta\mu}=0.
\end{equation}

In order to get the action of the teleparallel gravity, it is
convenient to define other two tensors:
\begin{equation}\label{contorsion}
K^{\mu\nu}_{~~\rho}=-\frac{1}{2}(T^{\mu\nu}_{~~\rho}-T^{\nu\mu}_{~~\rho}-T_{\rho}^{~\mu\nu}),
\end{equation}
and
\begin{equation}\label{S}
S_\rho^{~\mu\nu}=\frac{1}{2}(K^{\mu\nu}_{~~\rho}+\delta_\rho^{~\mu}T^{\theta\nu}_{~~\theta}-\delta_\rho^{~\nu}T^
{\theta\mu}_{~~\theta}).
\end{equation}
Then the torsion scalar as the teleparallel Lagrangian is defined by
\begin{equation}\label{T}
T=T^{\rho}_{~\mu\nu} S_{\rho}^{~\mu\nu}.
\end{equation}
The action of teleparallel gravity is expressed as
\begin{equation}
I=\frac{1}{16\pi G}\int {\rm d}^4 x~e\,T ,
\end{equation}
where $e=$det$(e^i_{\mu})=\sqrt{-g}$. Performing variation of the
action with respect to the vierbein, one can get the equations of
motion which are equivalent to the results of Einstein's theory of
general relativity.

Just as in the $f(R)$ theory, the generalized version of
teleparallel gravity could be obtained by extending the Lagrangian
density directly to a general function of the scalar torsion $T$ :
\begin{equation}
 I=\frac{1}{16\pi G}\int {\rm d}^4x~e\,f(T).
\end{equation}
This modification is expected to provide a natural way to understand
the cosmological observations, especially for the dark energy
phenomena, as a motivation. The variation of the action with respect
to vierbein leads to the following equations:
\begin{equation}\label{field eqn}
\begin{split}
{\big[}e^{-1}e^i_\mu\partial_\sigma(eS_i^{~\sigma\nu})-T^\rho_{~\sigma\mu}S_\rho^{~\nu\sigma}{\big]}f_T+
S_\mu^{~\rho\nu}\partial_\rho Tf_{TT}\\
-\frac{1}{4}\delta_\mu^{~\nu}f=4\pi GT_\mu^{~\nu} ,
\end{split}
\end{equation}
where $f_T$ and $f_{TT}$ represent the first and second order
derivative with respect to $T$ respectively, and
$S_i^{~\sigma\nu}=e_i^\rho S_\rho^{~\sigma\nu}$. ~$T_\mu^{~\nu}$ is
the energy-momentum tensor of the particular mater, with assuming
that matter couples to the metric in the standard form.

\section{The Validity of Birkhoff's Theorem}
\label{sec3} We consider the external vacuum gravitational field
solution of a spherically symmetric object. The spherically
symmetric metric can always be written in the following form:
\begin{equation}\label{metric}
{\rm d} s^2=A^2\!(t,r)~{\mathrm{d}}t^2-B^2\!(t,r)~{\mathrm{d}}r^2-r^2{\mathrm{d}}\theta^2-r^2 \sin^2\!\theta~{\mathrm{d}}\phi^2,
\end{equation}
where $A(t,r)$, $B(t,r)$ are arbitrary functions of the coordinates $t$ and $r$. The corresponding vierbein field directly reads
\begin{equation}
e^i_\mu={\rm diag}{\big(}A(t,r),~B(t,r),~r,~r\sin \theta{\big)},
\end{equation}
and the determinant of vierbein is $e=A(t,r)B(t,r)r^2\sin\theta$. Then the tensors defined in Eqs. (\ref{torsion},\ref{contorsion},\ref{S}) are determined, and the torsion scalar is given by
\begin{equation}\label{T2}
T=\frac{2{\big(}A(t,r)+2A^\prime\!(t,r)\,r{\big)}}{A(t,r)B^2(t,r)\,r^2},
\end{equation}
where a prime denotes the derivative with respect to $r$ while the
derivative with respect to $t$ will be denoted by a dot overhead,
according to which we will follow these convention throughout this
work.

For convenience, we introduce the  tensor $E_\mu^{~\nu}$ to stand
for the left hand side of Eq. (\ref{field eqn}), and the field
equation can be re-expressed then as
\begin{equation}
E_\mu^{~\nu}=4\pi GT_\mu^{~\nu}.
\end{equation}
Firstly, we consider a typically concrete $f(T)$ model in the
following form, which has been studied frequently in literature(e.g.
\cite{fT,fT1,fT2}):
\begin{equation}\label{fT}
f(T)=T+\alpha(-T)^n,
\end{equation}
where $\alpha$ and $n$ are real constants with arbitrary sign and
the negative sign for convenience in the second term can be absorbed
in the coefficient $\alpha$ if one likes. It will be shown later
that other more complicated forms of $f(T)$ models possess the same
conclusion of this form. Then we work out all the components of
$E_\mu^{~\nu}$, and find half of them are not vanishing, including
some quite complicated ones. The two components we used, fortunately
not very complex, are given by respectively
\begin{eqnarray}
E_1^{~0}&=&\frac{{\big[}\alpha n(-T)^{n-1}-1{\big]}\dot{B}}{A^2B\,r},\\
E_1^{~1}&=&\frac{1}{2B^2r^2} {\Big[}{\big(}(nB^2+1-2n)A+2(1-2n)rA^\prime{\big)} \nonumber \\
& &\cdot\alpha {\big(}-T{\big)}^{n-1}+2rA^\prime-{\big(}B^2-1{\big)}A {\Big]}.\label{E11}
\end{eqnarray}
Since the non-diagonal elements of energy-momentum tensor are equal
to zero, $E_1^{~0}$ always vanishes, restricting $B(t,r)$ to be only
the function of $r$, that is,
\begin{equation}\label{B}
B(t,r)=B(r).
\end{equation}
There is no density or pressure of matter in the external vacuum space, implying that $E_1^{~1}$ is also equal to zero. After some manipulation, Eq. (\ref{E11}) leads to
\begin{eqnarray}\label{constrain}
\frac{2^n \alpha\,n}{(B^2r)^{n-1}}{\bigg(}-\frac{2A^\prime}{A}&-&\frac{1}{r}{\bigg)}^{\!\!n-1}{\bigg[}
\frac{B^2-2+\frac{1}{n}}{r}-{\Big(}4-\frac{2}{n}{\Big)}\frac{A^\prime}{A}{\bigg]} \nonumber \\
& &+\frac{4A^\prime}{A}+\frac{2(1-B^2)}{r}=0.
\end{eqnarray}
It can be regarded as an algebraic equation of degree $n$ for $(A^\prime/A)$, with no analytical solutions generally if $n>2$. Nevertheless, for $B$ is independent of $t$, Eq. (\ref{constrain}) determines $(A^\prime/A)$ as an implicit function of $r$. So far as the solution exists, it could be expressed as
\begin{equation}\label{A'/A}
\frac{A^\prime(t,r)}{A(t,r)}=G{\big(}r,B(r),\alpha,n{\big)} \equiv g(r).
\end{equation}
The integration of the above equation with respect to the variable $r$ gives that
\begin{equation}\label{integral}
\ln A(t,r)=\int\!\!g(r){\rm d}r+C(t),
\end{equation}
where $C(t)$ as the integral constant, is an arbitrary function of $t$. Therefore the function $A(t,r)$ can be written as
\begin{equation}
A(t,r)={\rm e}^{\int\!\!g(r){\rm d}r}{\rm e}^{\scriptscriptstyle C(t)}.
\end{equation}
The factor ${\rm e}^{\scriptscriptstyle C(t)}$ can always be absorbed in the metric through a coordinate transformation $t\to t^\prime$, where $t^\prime$ is the new time coordinate defined as:
 \begin{equation}
 {\rm d}t^\prime={\rm e}^{\scriptscriptstyle C(t)} {\rm d}t.
 \end{equation}
Defining $\widetilde A(r)\equiv {\rm e}^{\int\!\!g(r){\rm d}r}$, the metric presented in Eq. (\ref{metric}) becomes
\begin{equation}\label{metric1}
{\rm d} s^2=\widetilde A^2(r)\,{\mathrm{d}}t^{\prime2}-B^2(r)\,{\mathrm{d}}r^2-r^2{\mathrm{d}}\theta^2-r^2 \sin^2\!\theta~
{\mathrm{d}}\phi^2.
\end{equation}
This is exactly a static metric which is required by the Birkhoff's
theorem validity. In the following we will discuss a general case
for the f(T) modified gravity model.

The $f(T)$ models investigated by several authors before take
variant forms, most of which are more complex than that we just
considered in (\ref{fT}) form. It is impracticable to express the
field equations of all these cases so specifically as in the
(\ref{constrain}). Nonetheless, without being concerned with the
particular forms of $f(T)$ models, we give the two field equations
in vacuum $E_1^{~0}=0$, $E_1^{~1}=0$ directly as
\begin{equation}
\frac{\dot Bf_T}{A^2Br}=0,\label{B2}
\end{equation}
\begin{equation}
\frac{B^2r^2f+{\big(}2B^2-4-8r(\frac{A^\prime}{A}){\big)}f_T}{4B^2r^2}=0.\label{E211}
\end{equation}
Eq.(\ref{B2}) also gives eq.(\ref{B}). Noting that $f$ and $f_T$ are
both functions of torsion $T$ which according to eq.(\ref{T2}) can
be re-expressed as
\begin{equation}
T=\frac{4}{B^2r}{\big(}\frac{1}{2r}+\frac{A^\prime}{A}{\big)},
\end{equation}
it is clear that the $A$ and $A^\prime$ in Eq. (\ref{E211}) only
present in the form of $(A^\prime/A)$, just as in the case of Eq.
(\ref{constrain}). Consequently, the relation (\ref{A'/A}) is
preserved, and the static metric so (\ref{metric1}) is obtained
again.

Note that the integral in (\ref{integral}) is performed over the
vacuum region, therefore the distribution and motion of the internal
source matter can not influence $\widetilde A(r)$ any way. The only
property of the source matter may appear in $\widetilde A(r)$ is the
total mass, or, more generally speaking, the total charge. We then
come to the conclusion that the spherically symmetric vacuum
solution of the $f(T)$ gravity must be static, and is independent of
the radial distribution and motion of the source matter, implying
that the Birkhoff's theorem still holds generally.

\section{Discussions and Conclusions}
\label{sec4} The Birkhoff's theorem is a significant feature of the
theory of general relativity, in analogy to the Gauss theorem in
electromagnetism or classical Newtonian gravity. It has been
confirmed to be valid in the Palatini formalism of $f(R)$ gravity,
while it no longer holds in the metric formalism
generally\cite{fR3,invalid fR2,invalid fR3}. Similar to the
origination of $f(R)$ theories, the $f(T)$ gravity is extended from
the teleparallel gravity which is equivalent to the theory of
general relativity. We prove in this brief report that the
Birkhoff's theorem holds in a typical $f(T)$ model with the form of
power law, and also holds in general $f(T)$ gravities. The validity
of this conclusion is independent of the concrete form of the $f(T)$
models. As a consequence, the significant inference in the theory of
general relativity, such as the in-existence of the monopole
gravitational wave, is also possessed in the $f(T)$ gravities.

The Birkhoff's theorem in the theory of general relativity leads to
a second inference which is often discussed in literature as that
the vacuum space-time inside a spherically distributed matter is
flat. It has few actual applications in astrophysics observations
indeed, since there is usually no vacuum cavity in celestial bodies.
Nevertheless, it is often considered as a help to understand the
fashion of the cosmic expansion \cite{harwit,peebles}. Because of
the globally isotropy and homogeneity properties of the observed
universe in large scales, the distribution of distant matter is
close to spherically symmetric about us. Though a great number of
galaxies at Hubble distance is moving away from us at the
relativistic speeds, the local effect can be neglected. That is to
say, in the background of the expanding
Friedmann-Robertson-Walker(FRW) universe, the ambient vacuum of a
spherically symmetric star or galaxy can be regarded as in the
static Schwarzschild geometry. It explains that the celestial bodies
move along Newtonian trajectories which are impervious to the cosmic
expansion.

It should be mentioned that the above inference, the second one,
which is not involved in the proof of this work directly, is
authentically valid only in Newtonian gravity and the theory of
general relativity without the cosmological constant. It generally
does not hold in modified gravity \cite{MOG}. It is just
approximately correct even in the standard $\rm{\Lambda}$CDM model
of cosmology. The non-vanishing cosmological constant $\rm{\Lambda}$
in Einstein's field equation is equivalent to a special matter with
the state parameter $\omega=-1$ everywhere, which breaks the vacuum
condition of the inside space. From the local point of view, with
the presence of cosmological constant $\rm{\Lambda}$, the vacuum
gravitational field is not described by the Schwarzschild solution
but by the Schwarzschild-de Sitter solution:
\begin{equation}
{\rm d}s^2={\Big(}1-\frac{2M}{r}-\frac{\Lambda}{3}r^2{\Big)}{\rm d}t^2-{\Big(}1-\frac{2M}{r}-\frac{\Lambda}{3}r^2{\Big)}^{-1}{\rm d}r^2-r^2{\rm d}\Omega^2
\end{equation}
where ${\rm d}\Omega^2={\rm d}\theta^2+\sin^2\theta{\rm d}\phi^2$.
The spacetime is not absolutely flat in the vacuum cavity where the
total mass or charge $M=0$. For the current value given by the
standard model of cosmology, $\Lambda \sim 10^{-52} \rm{m}^{-2}$,
the term $\frac{\Lambda}{3}r^2$ in the above metric can be neglected
in the scale of solar system where $r\sim 10^{13}\rm{m}$ when
compared with the first term O(1). Other topics involving the local
influences of the cosmic expansion have also been studied (e.g.
\cite{second1,second2}), showing that the effects are
inconsiderable. In the context of the $f(T)$ gravity, the validity
of the second inference is beyond the scope of this present paper,
for it involves the concrete solution of particular $f(T)$ models,
which still needs and is worth of further studies. We will leave it
for future work.

Besides the popular studies of the f(T) gravity models to cosmology
for mimicking the dark energy behaviors, there are also still lots
of interesting topics in its astrophysics applications, which in
some sense might be more practical and obviously is worthy of
further investigations.

\section*{Acknowledgement}
 We thank Prof. Lewis H Ryder for lots of interesting discussions on possible roles the torsion may play in
 gravity and cosmology physics during the
project over years.  This work is partly supported by Natural
Science Foundation of China under Grant Nos.11075078 and 10675062
and by the project of knowledge Innovation Program (PKIP) of Chinese
Academy of Sciences (CAS) under the grant No. KJCX2.YW.W10 through
the KITPC where we have initiated this present work.
%

%
\end{document}